\documentclass[12pt,preprint]{aastex}

\shorttitle{Massive outflows at high resolution}
\shortauthors{Beuther et al.}

\begin{document}

\title{Massive molecular outflows at high spatial
resolution}


\author{H. Beuther}
\affil{Harvard-Smithsonian Center for Astrophysics, 60 Garden Street, 
       Cambridge, MA 02138}
\email{hbeuther@cfa.harvard.edu}
\author{P. Schilke}
\affil{Max-Planck-Institut f\"ur Radioastronomie, Auf dem H\"ugel 69, 
       53121 Bonn, Germany}
\author{F. Gueth}
\affil{Institut de Radioastronomie Millimetrique (IRAM), 300 rue de la Piscine, 38406 Saint Martin d'Heres, France}

\begin{abstract}
We present high-spatial resolution Plateau de Bure Interferometer
CO(2--1) and SiO(2--1) observations of one intermediate-mass and one
high-mass star-forming region. The intermediate-mass region
IRAS\,20293+3952 exhibits four molecular outflows, one being as
collimated as the highly collimated jet-like outflows observed in
low-mass star formation sources. Furthermore, comparing the data with
additional infrared H$_2$ and cm observations we see indications that
the nearby ultracompact H{\sc ii} region triggers a shock wave
interacting with the outflow. The high-mass region IRAS\,19217+1651
exhibits a bipolar outflow as well and the region is dominated by the
central driving source. Adding two more sources from the literature,
we compare position-velocity diagrams of the intermediate- to
high-mass sources with previous studies in the low-mass regime. We
find similar kinematic signatures, some sources can be explained by
jet-driven outflows whereas other are better constrained by wind-driven
models. The data also allow to estimate accretion rates varying from a
few times $10^{-5}$\,M$_{\odot}$yr$^{-1}$ for the intermediate-mass
sources to a few times $10^{-4}$\,M$_{\odot}$yr$^{-1}$ for the
high-mass source, consistent with models explaining star formation of
all masses via accretion processes.
\end{abstract}

\keywords{accretion, accretion disks -- techniques: interferometric -- stars: formation -- ISM: jets and outflows -- ISM: individual (IRAS\,19217+1651; IRAS\,20293+3952)}

\section{Introduction}

Studies of massive molecular outflows have revealed many important
insights in the formation of massive stars over recent years.  Based
on the morphologies and energetics we can deduce physical processes
taking place at the inner center of the regions. Several single-dish
studies agree on the results that massive molecular outflows are
ubiquitous phenomena in massive star formation and that they are far
more massive and energetic than their low-mass counterparts (e.g.,
\citealt{shepherd1996a,ridge2001,zhang2001,beuther2002b}).

A point these studies disagree on is the degree of collimation of
massive outflows. Based on early studies by \citet{shepherd1996b} and
its follow-ups, it was believed that high-mass outflows tend to be
less collimated than low-mass flows. As the outflow collimation is
theoretically tightly connected with the accretion process, these
studies favored the idea that massive stars might form via different
physical processes, e.g., the coalescence of intermediate-mass
protostars at the very center of dense evolving cluster
\citep{bonnell1998,stahler2000,bally2002}.

However, recent observations by \citet{beuther2002b} show that the
previously claimed lower collimation of massive outflows is mostly an
observational artifact caused by the larger distances of the target
sources (on the average a few kpc) and too low spatial resolution of
most studies. Their data taken with the IRAM 30\,m telescope at a spatial
resolution of $11''$ are consistent with massive, bipolar outflows as
collimated as their low-mass counterparts. This implies that massive
stars can form in a qualitatively similar manner as low-mass stars,
just with accretion rates increased by orders of magnitude.

These latter observations are still based on single-dish observations,
and to substantiate the scenario a statistically significant number of
high-spatial-resolution interferometer studies of massive molecular
outflows is necessary. As a first step in that direction
\citet{beuther2002d} have observed the massive star-forming region
IRAS\,05358+3543 with the Plateau de Bure Interferometer (PdBI) in
CO(1--0), SiO(2--1) and H$^{13}$CO$^+$(1--0). They observed a massive
outflow from the central object of the evolving cluster which is
jet-like and highly collimated with a collimation degree of 10. This
is the upper end of collimation degrees observed for low-mass outflows
as well \citep{richer2000}. In addition to that collimated jet-like
structure, they observed at least two more outflows within the same
region.  In another source, IRAS\,19410+2336, the two outflows
observed at single-dish resolution split up at least into 7 separate
outflows when observed with interferometers
\citep{beuther2003a}. Similar results were observed toward G35.2 by
\citet{gibb2003}. One of the main conclusions of these studies is that
massive star-forming regions can appear confusing with single-dish
instruments, but that it is possible to disentangle the structures
with high enough spatial resolution into features well known from
low-mass star formation.

Contrary, other high-spatial-resolution studies of high-mass
star-forming regions indicate that massive outflows can also appear
morphologically and energetically different to their low-mass
counterparts (e.g., \citealt{shepherd1998,shepherd2003}). As the
high-spatial-resolution results are still based on poor statistical
grounds, we pursued massive outflow studies with the
PdBI\footnote{IRAM is supported by INSU/CNRS (France), MPG (Germany),
and IGN (Spain).}.  Here we present the results of two more regions~--
IRAS\,19217+1651 and IRAS\,20293+3952~-- observed at an angular
resolution as high as $1.8''$ in CO(2--1) and SiO(2--1). The two
sources are part of a large and well studied sample of 69 high-mass
protostellar objects at early evolutionary stages prior to producing
significant ultracompact H{\sc ii} (UCH{\sc ii}) regions
\citep{sridha,beuther2002a,beuther2002b,beuther2002c}. The two sources
were chosen because they combine different features of massive star
formation: IRAS\,19217+1651 has a luminosity of
$10^{4.9}$\,L$_{\odot}$ and shows a rather simple morphology with one
mm continuum source associated with cm emission and H$_2$O and class
II CH$_3$OH masers. Contrary, IRAS\,20293+3952 contains a small
UCH{\sc ii} region, contributing most of the bolometric luminosity
($10^{3.8}$\,L$_{\odot}$, \citealt{sridha}), and likely a cluster of
younger intermediate-mass sources triggering the molecular
outflows. While the single-dish outflow map of IRAS\,19217+1651 shows
a well-defined bipolar morphology, already the single-dish data of
IRAS\,20293+3952 show that we are dealing with multiple outflows in
that region \citep{beuther2002b}. Both sources cover a wide range of
characteristics from intermediate- to high-mass star formation, the
main source parameters are listed in Table \ref{sources}.

After describing the observations in \S \ref{obs}, we present the
observational results for both sources separately in \S \ref{obs_res}.
Then we discuss the results in the framework of massive star formation
and include literature data with special regard to the
position-velocity structure of massive outflows in \S
\ref{discussion}. Finally, \S \ref{conclusion} draws the conclusions,
summarizes the current stage of massive molecular outflow studies, and
outlines main topics to be tackled in the coming years.

\section{Observations}
\label{obs}

\subsection{Plateau de Bure Interferometer (PdBI)}

We observed IRAS\,19217+1632 and IRAS\,20293+3952 in different runs
from November 1999 to February 2002 with the Plateau de Bure
Interferometer at 1.3\,mm and at 3\,mm in the C and D configurations
with 4 antennas in 1999 and 5 antennas in 2002. The 1\,mm receivers
were tuned to the CO(2--1) line at 230.5\,GHz, and the 3\,mm receivers
covered the SiO(2--1) line at 86.85\,GHz.

The typical system temperatures at 1.3\,mm are about 300\,K and at
3\,mm about 135\,K. The phase noise was mostly below 20$^{\circ}$ and
always below 30$^{\circ}$. After smoothing the data, the final
velocity resolution in both lines is 1\,km\,s$^{-1}$, adequate to
sample the broad wing emission of the outflows. Atmospheric phase
correction based on the 1.3\,mm total power was applied. For continuum
measurements, we placed two 320\,MHz correlator units in each band to
cover the largest possible bandwidths. The primary beam at 1.3\,mm is
$22''$, and to cover both regions completely mosaics were
necessary. In IRAS\,19217+1651 the mosaic consisted of 5 fields
(offsets [$''$]: 19/33, 12/24, 6/14, 0/4, -5/-6) and in
IRAS\,20293+3952 of 8 fields (offsets [$''$]: 24/10, 12/7, 2/4, -10/1,
36/15, 32/3, 32/-9 32/-20) with respect to the phase reference centers
listed in Table \ref{sources}. The somewhat peculiar V-shaped mosaic
for IRAS\,20293+3952 was chosen based on the previous single-dish
observations \citep{beuther2002b}. Temporal fluctuations of the
amplitude and phase were calibrated with frequent observations of the
quasars 1923+210, 2032+107 and 2013+370. The amplitude scale was
derived from measurements of MWC349 and 3C345, and we estimate the
final flux density accuracy to be $\sim 15\%$. Synthesized beams at
the different wavelengths are listed in Table \ref{beams}.

\subsection{Short spacings with the IRAM 30\,m telescope}

To account for the missing short spacings and to recover the
line-flux, we also observed the source in CO(2--1) at $11''$
resolution with the IRAM 30\,m telescope in Summer 2002. The
observations were done remotely from the Max-Planck-Institut f\"ur
Radioastronomie (MPIfR) Bonn in the on-the-fly mode. Typical system
temperatures at 1.3\,mm were around 400\,K, the data were sampled in
$4''$ increments (Nyquist sampling is $\frac{\lambda}{2D} \sim 4.5''$)
and the velocity resolution was 0.1\,km\,s$^{-1}$.

The algorithm to derive visibilities from the single-dish data
corresponding to each pointing center is described by
\citet{gueth1996}. The single-dish and interferometer visibilities are
subsequently processed together. Relative weighting has been chosen to
minimize the negative side-lobes in the resulting dirty beam while
keeping the highest angular resolution possible. Images were produced
using natural weighting, then a CLEAN-based deconvolution of the
mosaic was performed. Synthesized beams of the merged data are listed
in Table \ref{beams}.

It should be noted that only the CO(2--1) data have been complemented
by short spacing data from the 30 m telescope whereas we did not get
these data at 3\,mm. Nevertheless, at 3\,mm the short-spacings problem
is less severe because the interferometer samples larger regions at
lower frequencies.

\section{Observational results}
\label{obs_res}

For both sources we detect bipolar outflows in CO(2--1) and SiO(2--1)
(Figs. \ref{19217_co} \& \ref{20293_co}). While IRAS\,19217+1651 is
dominated by one extremely energetic outflow, IRAS\,20293+3952
exhibits one collimated jet-like outflow and at least three more
outflows. Before discussing their implications for massive star
formation we present each star-forming region separately.

\subsection{IRAS\,19217+1651}
\label{19217}

\subsubsection{Millimeter Continuum}
\label{cont_19217}

Even at the highest spatial resolution of $\sim 1.5''$ the mm
continuum emission of IRAS\,19217+1651 remains single-peaked and does
not split up into multiple sub-sources (Fig. \ref{19217_cont}).  The
millimeter continuum fluxes are given in Table
\ref{continuum}. Comparing the 1.3\,mm flux obtained with the PdBI and
the 1.2\,mm single-dish fluxes \citep{beuther2002a} we estimate that
about $80\%$ of the total continuum flux is filtered out by the
interferometer. Compact cm continuum emission and 22\,GHz H$_2$O and
6.7\,GHz Class II CH$_3$OH maser emission observed with the VLA and
ATCA \citep{sridha,beuther2002c} peak at the mm continuum source
\footnote{\citet{beuther2002c} presented a similar image with the cm
source and one H$_2$O maser feature being $\sim 5''$ offset to the
east. Unfortunately, the astrometry in their image was wrong, here we
present the correct positions.}. It has to be taken into account that
IRAS19217+1651 is five times further away than IRAS\,20293+3952, and
thus we cannot resolve as much structure. Nevertheless, the spatial
coincidence of mm/cm continuum emission and the two maser species
suggests that the region is dominated by one massive evolving protostar
at the cluster center.

Assuming optically thin dust emission at mm wavelength, we calculate
the mass and peak column density using the 1.3\,mm data following the
procedure outlined for the single-dish dust continuum data by
\citet{beuther2002a}. Recent studies indicate that the dust opacity
index $\beta$ could be lower than the canonical value 2 at the core
centers of massive star-forming regions (\citealt{goldsmith1997} and
references therein; \citealt{beuther2004b,beuther2004e,kumar2003}).
Unfortunately, we cannot properly differentiate between the dust and
free-free contributions in the mm regime toward IRAS\,19217+1615, and
thus not derive $\beta$ explicitly for this source. Based on the other
studies we set $\beta$ to 1. We use a dust temperature of 38\,K as
derived by SED fits to the IRAS data \citep{sridha}. As discussed by
\citet{beuther2002a}, the errors of the estimated masses and column
densities are dominated by systematics like the exact knowledge of
$\beta$ or the temperature. For example, reducing $\beta$ from the
canonical value of 2 to 1 lowers the estimated mass by about one order
of magnitude. We estimate the masses and column densities to be
correct within a factor 5-10. The total gas mass of the central core
observed at 1.3\,mm with the PdBI is around 220\,M$_{\odot}$, and the
peak column density of the order a few times $10^{23}$ cm$^{-2}$
corresponds to a visual extinction $A_{\rm{v}} =
N_{\rm{H}}/2\times10^{21} \sim 600$ (Table \ref{continuum}).

\subsubsection{The molecular outflow}
\label{outflow_19217}

Figure \ref{19217_co} presents the merged PdBI+30m CO(2--1) outflow
image obtained for IRAS\,19217+1651. We observe a bipolar outflow
emanating from the mm core with the main blue emission to the
south-west and the main red emission to the north-east. The gas with
the highest velocities ($\pm 30$\,km\,s$^{-1}$) is located at the core
center, however we find high velocity gas offset from the core as well
(see below, Figure \ref{pos_velo}). The overall collimation of the
outflow is pretty high with a collimation degree $\sim 3$ (length of
the outflow divided by its width). The morphology of the blue outflow
wing resembles a cone-like structure. The red wing to the north-east
shows an elongated feature with a P.A. of $40^{\circ}$ with respect to
the main outflow axis. The morphology of the red wing is quite
different compared with the blue wing, and one can get the impression
that in this area might be a second outflow in the north. However, the mass
contained in the red wing is very high ($\sim 50$\,M$_{\odot}$, Table
\ref{outflows}), and it is difficult to imagine such a massive and
energetic outflow without a mm continuum source as the counterpart
(the $3\sigma$ rms of 7.5\,mJy corresponds to a mass sensitivity of
4.3\,M$_{\odot}$). Therefore, we conclude that the red and blue wing
emission is part of the same outflow emanating from the massive mm
core. The different morphologies to the north and south are likely
attributed to environmental differences.

The SiO(2--1) emission shown in Figure \ref{19217_co} exhibits a
similar outflow morphology as the CO observations but missing the
broad lower intensity outflow emission depicted in CO. As SiO is
mainly a shock tracer \citep{schilke1997a} it is likely that SiO is
not excited in the outer regions of the outflow. However, it should be
mentioned that this difference might also be an observational artifact
because we do not have the short spacings data for SiO and thus the
larger scale SiO outflow emission could be filtered out by the
interferometer. Furthermore, due to the larger primary beam of the
PdBI at 3\,mm compared to 1\,mm ($59''$ and $22''$, respectively) we
observe a larger field in SiO and detect an additional bipolar
structure in the west not covered by the CO data. We do not detect a
mm continuum source there (the 3\,mm continuum covers the same field).
Nevertheless, it is likely that a mm continuum source is simply too
weak and not detected due to insufficient signal to noise. Adopting
the same assumptions outlined in \S \ref{cont_19217} for the 3\,mm
continuum data, the $3\sigma$\,rms of 2.4\,mJy corresponds to a mass
sensitivity of $\sim 30$\,M$_{\odot}$.


\subsection{IRAS\,20293+3952}
\label{20293}

\subsubsection{Millimeter Continuum}
\label{cont_20293}

Figure \ref{20293_cont} presents the mm continuum data for
IRAS\,20293+3952 and additionally the cm and H$_2$O maser emission in
that region. Obviously, the overall picture is different from
IRAS\,19217+1651. We find three mm continuum sources with an H$_2$O
maser associated with the strongest of them. Furthermore, offset from
the mm emission there is a resolved cm source indicating a more
evolved UCH{\sc ii} region. We do not detect any mm continuum emission
at the position of the UCH{\sc ii} region down to the $3\sigma$ rms
sensitivity limit of 13.5\,mJy, corresponding to a mass sensitivity of
0.2\,M$_{\odot}$ assuming optically thin dust emission. Instead of one
source dominating the whole region (\S \ref{cont_19217})
IRAS\,20293+3952 exhibits four sources possibly interacting with each
other. Comparing the interferometric and single-dish fluxes
\citep{beuther2002b}, nearly $90\%$ of the total flux is filtered out
in IRAS\,20293+3952, even more than in IRAS\,19217+1651.

Assuming optically thin dust emission with a dust temperature of 56\,K
\citep{sridha} using again the dust opacity index $\beta=1$ (\S
\ref{cont_19217}) we derive masses and column densities for the three
mm clumps.  The masses of each clump listed in Table \ref{continuum}
are about two orders of magnitude below the value derived for
IRAS\,19217+1651 whereas the beam averaged column densities are of the
same order, only lower by factors of 2-6.

The clump masses between 1 and 3\,M$_{\odot}$ appear low regarding the
overall luminosity of the region of $10^{3.8}$\,L$_{\odot}$.  However,
the luminosity is measured with the large IRAS beam and thus comprises
also the nearby UCH{\sc ii} region. Assuming the cm emission to be
optically thin, \citet{sridha} estimated the stellar luminosity of its
central source to be close to the infrared-derived value. Therefore,
while it is possible that we partly underestimate the mass of
the dust cores in IRAS\,20293+3952, it seems obvious that the dominant
luminosity source is the UCH{\sc ii} region. The mm sources nearby,
which trigger all the spectacular outflows, form a kind of secondary
cluster of intermediate-mass sources.

\subsubsection{Four molecular outflows}
\label{outflow_20293}

As the millimeter continuum, the outflow emission in IRAS\,20293+3952
is more complex than in IRAS\,19217+1651 as well. The CO velocity
spread down to zero intensity with $\Delta v \sim 92$\,km\,s$^{-1}$ in
IRAS\,20293+3952 is also larger compared to $\Delta v \sim
64$\,km\,s$^{-1}$ in IRAS\,19217+1651.  Figure \ref{20293_co} shows
three images of the CO red and blue outflow emission (extreme outflow
velocities, moderate outflow velocities, and all outflow velocities),
it also sketches the four outflows identified in that region. Figure
\ref{20293_sio} shows the SiO(2--1) emission of the region. As the
region is complex not all identifications are unambiguous and some
features can belong to one or the other outflow, or even outflows
which we do not identify yet. Our outflow identifications are based on
the CO and SiO morphology and the assumption that there are no other
outflow-powering sources than mm1 to mm3.

{\it Outflow (A):} The most prominent feature in Figure \ref{20293_co}
is the collimated jet-like outflow emanating from mm1 in
south-west--north-eastern direction. Especially interesting is the red
wing with its extreme collimation extending about 1\,pc in length. The
collimation degree of this red wing is $\sim 8$ as high as known
values for the most collimated low-mass outflows
\citep{richer2000}. Figure \ref{red_20293} shows a close-up of
moderate and extreme velocities for the red emission stressing that
the higher-velocity gas is more collimated than the lower-velocity
gas. The ratio of the projected width perpendicular to the outflow
direction of the moderate-velocity gas versus the extreme-velocity gas
is $\ge 2$. This morphology resembles the observations of low-mass
outflows \citep{bachiller1996}. It is unlikely that the
moderate-velocity gas is strongly confused by the ambient gas because
the chosen velocity interval [13,25]\,km\,s$^{-1}$ for the
moderate-velocity component is still significantly offset from the
velocity of rest $v_{\rm{LSR}}=6.5$\,km\,s$^{-1}$. Furthermore, Figure
\ref{pos_velo} presents a position velocity diagram of outflow (A),
and we find high-velocity gas at the outflow center as well as at the
very end of the red wing. The morphology of the blue wing is less
clear which might be partly due to the smaller extend of the mosaic in
that direction (\S \ref{obs}). The blue feature highlighted with
the dashed arrow in Fig. \ref{20293_co} could be part of a larger cone
of the blue wing of (A), but it is also possible that there is another
outflow consisting of this blue feature and maybe the northern red
feature we so far associate with outflow (D).

{\it Outflow (B):} A second outflow emanates from mm1 in
south-eastern direction. The blue CO emission is only observed at a
distance of $\sim 20''$ but SiO is observed right to mm1. While the
blue emission is strong we do not observe red emission to the
north. Partly, this should be due to the small extend of the PdBI
mosaic in that direction (\S \ref{obs}).

{\it Outflow (C):} We observe a third small outflow emanating from mm3
in east-west direction.

{\it Outflow (D):} The fourth outflow (D) likely emanates from mm2
and is more or less parallel to outflow B. Again, we see strong blue emission
toward the south-east but less red emission toward the north-west. As
already mentioned in the context of outflow (A), the red feature in
the north could also be part of still another outflow emanating from
mm1 with the blue counterpart shown in dashed contours. \\

{\it Shocked H$_2$ emission:} Additionally interesting is a comparison
of the outflow data with shocked H$_2$ emission at 2.12\,$\mu$m. The
near-infrared H$_2$ data were taken with the Omega Prime camera on the
Calar Alto telescope within the observations of a large sample of
massive star-forming regions, the basic data reduction is described in
a separate paper by Stanke et al. (in prep.). Figure \ref{20293_h2}
presents an overlay of the H$_2$ emission with the CO outflow data and
the cm source outlining the UCH{\sc ii} region. We can distinguish a
few regions of prominent H$_2$ emission: first, we find H$_2$ features
associated with the blue CO emission of outflows (A) and
(B). Furthermore, we clearly identify ring-like H$_2$ emission around
the UCH{\sc ii} region. Judging from the morphology, the red CO
outflow (A) bends partly around that ring-like structure. In addition,
there are strong H$_2$ emission knots between mm2 and mm3 which could
be associated with outflows (C) and (D), but which may also be part of
the ring-like H$_2$ emission.

\section{Discussion}
\label{discussion}

\subsection{Molecular outflows}

\subsubsection{Masses and energetics}

The CO(2--1) data allow to estimate the masses and energetics of the
different outflows in both sources. However, for IRAS\,20293+3952 it
is not always clear whether emission belongs to one or the other
outflow. For example, the emission right at the center of mm1 can be
part of the outflows (A) as well as (B). Most likely, both outflows
contribute to the observed emission. Similarly, the emission between
mm2 and mm3 could be part of the outflows (C) and (D). In the
following calculations these regions of overlap are always attributed
to both contributing outflows.

We calculate opacity-corrected H$_2$ column densities in both outflows
following the approach outlined in \citet{beuther2002b}. The average
temperature in the outflows is set to 30\,K and the average line
opacity in the outflow wings to $\tau (^{13}$CO~$2-1)=0.1$ (based on
observations by \citealt{choi1993}). According to \citet{cabrit1990}
derived masses are accurate to a factor 2 to 4, whereas the accuracy
of dynamical parameters are lower, at about a factor 10. The derived
masses and energetic parameters for the outflows in both sources are
presented in Table \ref{outflows}.

A comparison of the outflow mass $M_{\rm{out}}=75$\,M$_{\rm{\odot}}$
in IRAS\,19217+1651 with the single-dish observations derived value of
108\,M$_{\rm{\odot}}$ \citep{beuther2002b} shows that both values
agree within 25$\%$. This gives confidence that the merging process of
the PdBI data with the single-dish observations worked reasonably well
and we recovered most of the outflow emission. Such a comparison 
is more difficult for IRAS\,20293+3952 because the single-dish
data did not resolve the multiple outflows.

The values presented in Table \ref{outflows} confirm that
IRAS\,19217+1651 powers a very massive and energetic molecular outflow
on pc scales. The derived outflow rate $\dot{M}_{\rm{out}}$ is of the order
$10^{-3}$\,M$_{\rm{\odot}}$yr$^{-1}$. Under the assumption of momentum
driven outflows this leads to an estimate of the accretion rate of the
order a few times $10^{-4}$\,M$_{\rm{\odot}}$yr$^{-1}$ (for details on the
assumptions see \citealt{beuther2002b}). Accretion rates of that order
are high enough to overcome the radiation pressure of a forming star
and build the most massive stars via accretion processes (e.g.,
\citealt{norberg2000,mckee2003}).

The outflow parameters for IRAS\,20293+3952 are all about one order of
magnitude below the values for IRAS\,19217+1651. They are higher than
typical masses and energetics in the low-mass regime
\citep{richer2000} but below often observed parameters for massive
outflows \citep{shepherd1996b,ridge2001,beuther2002b,gibb2003}. The
outflows emanate from a cluster of intermediate-mass protostars right
in the vicinity of a more evolved UCH{\sc ii} region. The data show
that the UCH{\sc ii} region is too evolved to trigger any collimated
outflow. Thus, although most of the bolometric luminosity stems from
the UCH{\sc ii} region, nearly all the mechanical force $F_{\rm{m}}$
and mechanical luminosity $L_{\rm{m}}$ (Table \ref{outflows}) is due
to the outflows from the intermediate-mass cluster. Based on the
outflow rate $\dot{M}_{\rm{out}}$ we can again estimate accretion rates
between $10^{-5}$ and $10^{-4}$\,M$_{\rm{\odot}}$yr$^{-1}$, right between
typical values for low- and high-mass star-forming regions.

\subsubsection{Dynamical interactions in IRAS\,20293+3952}

The presence of four molecular outflows emanating from three mm
continuum sources, and one UCH{\sc ii} region within less than 1\,pc
projected on the plane of the sky reveals multiple regions of
interaction. Obviously, the CO and SiO emission around the mm sources
is caused by different interacting molecular outflows, but the UCH{\sc
ii} seems to affect the outflows as well. Assuming that the ring-like
shocked H$_2$ emission is caused by the star powering the central
UCH{\sc ii} region, its sphere of influence extends as far as outflow
(A) and the mm sources mm2 and mm3. The H$_2$ emission between mm2 and
mm3 might thus not be just due to the outflows (C) and (D) but there
may also be contributions from the UCH{\sc ii} region. With the data
so far, we cannot unambiguously conclude whether the spatial
association of mm2 and mm3 with the H$_2$ ring is simple coincidence
or whether the formation of mm2 and mm3 might even be associated with
some triggering mechanism from the UCH{\sc ii} region.

Even more intriguing is the spatial bending of the red wing of outflow
(A) right north of the UCH{\sc ii} region. Figure \ref{20293_h2} shows
that this bending follows largely the ring-like H$_2$ feature in that
region. This could be just due to projection effects, but it is also
possible that an expanding pressure wave from the UCH{\sc ii} region
intercepts with outflow (A) and pushes its gas slightly to the
north. The radius of the H$_2$ shell is about $7.7''$ corresponding to
15000\,AU at a distance of 2\,kpc.

\subsection{Position-velocity diagrams of massive outflows}

Position-velocity (p-v) diagrams are often used as a tool to
understand the driving mechanisms of outflows (e.g.,
\citealt{smith1997,downes1999,lee2001,lee2002}). In addition to p-v
diagrams of outflows presented in this paper, Figure \ref{pos_velo}
shows p-v diagrams of two other massive outflow sources taken from the
same initial source sample: IRAS\,23033+5951 observed
in CO(1--0) with BIMA by Wyrowski et al. (in prep.) and
IRAS\,20126+4104 observed in CO(2--1) with the IRAM 30\,m by Lebron et
al. (in prep.). For details on the outflow characteristics, we refer
to the corresponding papers, here we are only interested in their p-v
diagrams.

The range of outflow masses for these four outflows is broad, from
intermediate masses in IRAS\,20293 to very high masses in IRAS\,23033
($M_{\rm{out}}=119$\,M$_{\odot}$, Wyrowski et al., in prep.). The
bolometric luminosities also vary by orders of magnitude: outflow (A)
in IRAS\,20293 is driven by an intermediate-mass protostar, and the
bolometric luminosities of IRAS\,20126, IRAS\,23033 and IRAS\,19217
are $10^{3.9}$, $10^{4.0}$ and $10^{4.9}$\,L$_{\odot}$,
respectively. Admittedly, the statistical number of presented outflows
is low, but nevertheless we cover a broad range of luminosities and
masses and can compare these data with theories and previous studies
in the low-mass regime.

We observe mainly two features in the p-v space: first of all,
high-velocity gas is detected at the outflow centers in IRAS\,19217
and IRAS\,20293. There is some high-velocity gas in IRAS\,23033 near
the center as well, but the red and blue is shifted spatially to the
other side of the outflow center compared with the larger scale flow,
thus this high-velocity gas might be due to a second outflow spatially
just barely resolved (Wyrowski et al., in prep.). IRAS\,20126 does not
show any high-velocity gas near the outflow center. Secondly, we
observe high-velocity gas at some distances from the core center:
IRAS\,23033 exhibits the so-called Hubble-law in the blue wing, i.e.,
a velocity increase with distance from the core center. In
IRAS\,20293, the velocity also increases gradually with distance in
the red wing, and the emission feature at the very end of the
collimated outflow shows emission at all velocities again. In
IRAS\,20126, we see a gradual velocity increase and then decrease
again with distance from the core center, this is the only source
where the p-v diagram is rather symmetric. IRAS\,19217 shows some
high-velocity features at distances from the center as well, but there
is no real symmetry with respect to the core center.

We compare our data with high-spatial-resolution observations of
low-mass outflows by \citet{lee2000,lee2002}. They observed 10
low-mass sources and found mainly two kinematic features in the p-v
diagrams: parabolic structures originating at the driving source and
convex spur structures with high-velocity gas near H$_2$ bow
shocks. While the parabolic structures can be explained by wind-driven
models the spur structures are attributed to jet-driven bow-shock
models \citep{lee2000,lee2001,lee2002}. While some outflows show clear
signatures of one or the other model, there are also a few sources
which exhibit signatures of both. \citet{lee2002} propose that a
combination of a jet- and wind-driven model might explain all features
in a more consistent way. In their first published sample, they see
nearly no central high-velocity gas \citep{lee2000}, whereas in later
published sources some central high-velocity gas is observable
\citep{lee2002}. High-velocity features at the core centers are likely
due to jets but they can also be mimicked by highly inclined winds
(Fig. 10, \citealt{lee2001}).

Transferring the low-mass and simulation results to our data, we find
clear spur and Hubble-law jet-signatures in IRAS\,20293 and
IRAS\,23033, whereas IRAS\,20126 can be explained by a wind-driven
model (e.g., compare with RNO\,91, Fig.\,10, in
\citealt{lee2000}). IRAS\,19217 is a less clear-cut case because we
find the jet-indicating high-velocity gas at the core center but also
features further outside which resemble more the parabolic structures
indicative of wind-driven models. Morphologically, the
IRAS\,19217 outflow is also somewhat intermediate between a collimate
jet-like outflow and cavity-like features indicative rather of a wind
(Figure \ref{19217_co}). Likely, both mechanisms contribute to the
observed outflow in IRAS\,19217.

To summarize, our intermediate- to high-mass outflow data show
kinematic signatures which can be explained by jet- and/or wind-driven
models. We do not find any striking difference to low-mass
position-velocity diagrams. Similar to their low-mass counterparts, no
single model is yet capable to explain all observations consistently.

\section{Conclusions}
\label{conclusion}

The presented analysis of high-spatial-resolution observations of
intermediate- to high-mass molecular outflows indicates that the
outflow morphologies/kinematics and thus their driving mechanisms do
not vary significantly compared to their low-mass counterparts. The
higher the source luminosity the more energetic the outflows, but the
qualitative signatures are similar. These observations indicate that
similar driving mechanisms can be responsible for outflows of all
masses.

We find an extremely collimated jet-like outflow emanating from the
intermediate-mass source mm1 in IRAS\,20293+3952. This outflow shows the
highest degree of collimation at highest velocities and slightly lower
collimation at lower velocities, similar to low-mass sources
\citep{bachiller1996}. The whole region IRAS\,20293+3952 shows many
signs of dynamical interactions, not only between the different
outflows (at least four) but there are also indications of a shock wave
from the nearby UCH{\sc ii} region interacting with the collimated
outflow. While the UCH{\sc ii} region is the main source of luminosity
in IRAS\,20293, most of the mechanical force stems from the outflows
of the intermediate-mass sources.

The high-mass source IRAS\,19217+1651 shows a nice bipolar outflow,
slightly less collimated than the outflow (A) in IRAS\,20293, but
still comparable to many low-mass flows. This region is strongly
dominated by the central core which exhibits mm/cm continuum emission
as well as H$_2$O and CH$_3$OH maser emission.

Position-velocity diagrams of molecular outflows from intermediate to
high masses show similar signatures as known for low-mass
outflows. Some sources are better explained by jet-driven outflows
whereas others seem to be due rather to wind-driven
outflows. IRAS\,19217 exhibits signatures of both. The proposal from
\citet{lee2002} that a combination of both driving mechanisms
can explain all outflows consistently also holds for our sample.

Estimated accretion rates are of the order a few times
$10^{-5}$\,M$_{\odot}$yr$^{-1}$ for the intermediate-mass sources in
IRAS\,20293 and a few times $10^{-5}$\,M$_{\odot}$yr$^{-1}$ for the
high-mass source IRAS\,19217, consistent with models forming stars of
all masses via accretion (e.g, \citealt{norberg2000,mckee2003}).

The data presented in this paper further support the idea that
massive stars form via similar accretion-based processes as their
low-mass counterparts. The main difference appears to be their
clustered mode of formation and increasing accretion rates and
energetics with increasing stellar mass and luminosity. However,
investigations of the most massive stars is just beginning, and
studies like this one so far rarely exceeded sources with luminosities
$>10^5$\,L$_{\odot}$. This is to a large degree due to the fact that
there simply do not exist many sources with far higher luminosity
which are in a state of evolution prior or at the very beginning to
form a significant UCH{\sc ii} region. Therefore, on the one hand we
have to extend massive star formation research significantly to even
higher luminosities to confirm the present results or to identify
possible differences in that regime. On the other hand, the constrains
set on the massive star-forming processes are yet mostly indirect,
e.g., observing molecular outflows on large scales and inferring the
processes likely taking place at the cluster centers. As the spatial
resolution and sensitivity of (sub-)mm interferometers increase
steadily, it is now necessary to really study the cluster centers and
try to resolve the relevant processes in more direct ways. For
example, massive disk which are crucial to explain the observed
outflows need to be properly identified, resolved and studied to
manifest its physical conditions. Furthermore, the strong radiation of
the massive protostars significantly changes the chemistry of those
central regions. The broad bandwidth and high spatial resolution of
current and future (sub-)mm interferometers (SMA, PdBI, CARMA, and
further on ALMA) will shed light on many such processes.

\acknowledgments We like to say thank you very much to Thomas Stanke,
Friedrich Wyrowski and Mayra Lebron for providing the H$_2$ data and
two of the p-v diagrams prior to publication.  We also thank
G. Paubert from the IRAM 30\,m telescope for help with the single-dish
data. Furthermore, we thank the referee for helpful comments
improving the presentation of the paper. H.B. acknowledges financial
support by the Emmy-Noether-Program of the Deutsche
Forschungsgemeinschaft (DFG, grant BE2578/1).


\clearpage


\begin{deluxetable}{lrrrrrrrr}
\tablecaption{Source characteristics \label{sources}}
\tablewidth{0pt}
\tablehead{
Source & R.A. & Dec. & $v_{\rm{LSR}}$& $D$ & $L$ & $M$ & $M_{\rm{out}}$ & $E_{\rm{out}}$ \\
       & J2000& J2000& km\,s$^{-1}$  & kpc & L$_{\odot}$ & M$_{\odot}$ & M$_{\odot}$ & erg  
}
\startdata
19217+1632 & 19:23:58.78 & 16:57:36.52 & 3.5 & 10.5 & $10^{4.9}$ & 9500 & 108 & $3.6\,10^{47}$ \\
20293+3952 & 20:31:10.70 & 40:03:09.98 & 6.3 & 2.0  & $10^{3.8}$ & 460  & 9   & $7.8\,10^{46}$  
\enddata
\tablecomments{\footnotesize{The source parameters are taken from
\citet{sridha,beuther2002a,beuther2002b}: the velocity of rest
$v_{\rm{LSR}}$, the distance $D$, the luminosity $L$, the core mass
$M$, the outflow mass $M_{\rm{out}}$ and the outflow energy
$E_{\rm{out}}$}}
\end{deluxetable}

\begin{deluxetable}{lrrrr}
\tablecaption{Synthesized beams \label{beams}}
\tablewidth{0pt}
\tablehead{
\colhead{Source} & \colhead{Wavel.} & Obs & \colhead{Beam} & \colhead{lin. res.} \\
                 & \colhead{mm}   &     & \colhead{$''$(P.A.)} & \colhead{AU} 
}
\startdata
19217+1632 & 1.3 cont & PdBI & $1.58\times 1.43$ (32$^{\circ}$) & $\sim 16600 \times 15000$ \\
19217+1632 & 1.3 line & PdBI+30m & $1.93\times 1.69$ (51$^{\circ}$) & $\sim 20300 \times 17700$\\
19217+1632 & 3 cont   & PdBI & $4.93\times 3.78$ (58$^{\circ}$) & $\sim 51800 \times 39700$\\
19217+1632 & 3 line   & PdBI & $6.05\times 4.91$ (71$^{\circ}$) & $\sim 63500 \times 51600$\\ 
\hline
20293+3952 & 1.3 cont & PdBI & $1.91\times 1.75$ ($-101^{\circ}$) & $\sim 3800 \times 3500$\\
20293+3952 & 1.3 line & PdBI+30m & $1.96\times 1.79$ (84$^{\circ}$) & $\sim 3900 \times 3600$\\
20293+3952 & 3 cont   & PdBI & $5.10\times 4.31$ (50$^{\circ}$) & $\sim 10200 \times 8600$\\
20293+3952 & 3 line   & PdBI & $5.10\times 4.36$ (49$^{\circ}$) & $\sim 10200 \times 8700$
\enddata
\end{deluxetable}

\begin{deluxetable}{lrrrrrrr}
\tablecaption{Millimeter continuum data\label{continuum}}
\tablewidth{0pt}
\tablehead{
\colhead{Source} & \colhead{$\#$} & \colhead{$S_{\rm{peak}}^{\rm{3mm}}$} & \colhead{$S_{\rm{int}}^{\rm{3mm}}$} & \colhead{$S_{\rm{peak}}^{\rm{1mm}}$} & \colhead{$S_{\rm{int}}^{\rm{1mm}}$} & \colhead{$M$} & \colhead{$N$} \\
\colhead{} & \colhead{} & \colhead{$\frac{\rm{mJy}}{\rm{beam}}$} & \colhead{mJy} & \colhead{$\frac{\rm{mJy}}{\rm{beam}}$} & \colhead{mJy} & \colhead{M$_{\odot}$} & \colhead{cm$^{-2}$}
}
\startdata
19217+1632 &   & 56 & 68         & 100& 379 & 216 & 6\,$10^{23}$\\
20293+3952 & 1 & 13 & 19         & 95 & 201 & 3  & 2\,$10^{23}$\\
20293+3952 & 2 &  7$^a$ & 12$^a$ & 47 & 92  & 1  & 1\,$10^{23}$\\
20293+3952 & 3 & --     & --     & 37 & 76  & 1  & 1\,$10^{23}$ 
\enddata
\tablenotetext{a}{\footnotesize{Sources mm2 and mm3 are unresolved at 3\,mm.}}
\end{deluxetable}

\begin{deluxetable}{lrrrrrrrrrrr}
\tablecaption{Outflow parameter \label{outflows}}
\tablewidth{0pt}
\tablehead{
\colhead{source} & \colhead{$M_{\rm{blue}}$} & \colhead{$M_{\rm{red}}$} & \colhead{$M_{\rm{out}}$} & \colhead{$p$} & \colhead{$E$} & \colhead{size} & \colhead{$t_{\rm{dyn}}$} & \colhead{$\dot{M}_{\rm{out}}$} & \colhead{$F_{\rm{m}}$} & \colhead{$L_{\rm{m}}$}\\
\colhead{} & \colhead{M$_{\odot}$} & \colhead{M$_{\odot}$} & \colhead{M$_{\odot}$} & \colhead{$\frac{\rm{M_{\odot}km}}{\rm{s}}$} & \colhead{$10^{46}$erg} & \colhead{pc} & \colhead{yr} & \colhead{$\frac{\rm{M_{\odot}}}{\rm{yr}}$} & \colhead{$\frac{\rm{M_{\odot}km}}{\rm{s\,\,yr}}$} & \colhead{L$_{\odot}$}
}
\startdata
19217+1652    & 24.4 & 50.4 & 74.8 & 2210 & 68  & 1.6  & 48200 & 1.5e-3 & 4.6e-2 & 116\\
20293+3952(A) & 1.3  & 0.7  & 2.0  &  90  & 4.1 & 0.20 &  4300 & 4.5e-4 & 2.1e-2 & 79 \\ 
20293+3952(B) & 1.0  & 0.0  & 1.0  &  46  & 2.1 & 0.17 &  3700 & 2.7e-4 & 1.2e-2 & 46 \\ 
20293+3952(C) & 0.2  & 0.1  & 0.3  &   9  & 0.2 & 0.06 &  2100 & 1.5e-4 & 4.1e-3 &  9 \\ 
20293+3952(D) & 0.8  & 0.1  & 0.9  &  22  & 0.6 & 0.31 & 12600 & 6.8e-5 & 1.7e-3 &  4 
\enddata
\tablecomments{\footnotesize{The listed parameters are the outflow masses in the blue wing $M_{\rm{blue}}$, the red wing $M_{\rm{red}}$ and the total mass $M_{\rm{out}}$, the momentum $p$, the energy $E$, the size of the flows, their dynamical timescale $t_{\rm{dyn}}$, the outflow rate $\dot{M}_{\rm{out}}$ and their mechanical forces $F_{\rm{m}}$ and luminosities $L_{\rm{m}}$.}}
\end{deluxetable}

\clearpage


\begin{figure}[htb] 
\caption{Continuum emission in IRAS\,19217+1651: the grey-scale
presents the 3\,mm and the thin contours the 1.3\,mm continuum. The
thick contours show the 3.6\,cm continuum emission. The symbols point
at the maser as labeled in the top right. The large and small beams at
the bottom right are from the 3\,mm and 1.3\,mm observations,
respectively. The mm data are contoured from 20 to 90\% (10\% steps)
from the peak emission (Table \ref{continuum}). The cm observations
are contoured from 30 to 90\% (10\% steps) from the peak emission of
19.6\,mJy/beam. \label{19217_cont}}
\end{figure}

\begin{figure}[htb] 
\caption{CO(2--1) (left panel) and SiO(2--1) (right panel)
observations of IRAS\,19217+1632: The grey scale with thick contours
presents the 1.3\,mm continuum emission contoured from 15 to 95\%
(10\% steps) from the peak emission (Table \ref{continuum}). Full and
dotted contours show the red and blue shifted emission in both
species, respectively. The velocity ranges for the CO(2--1) data are:
blue v=[-35,-3] km\,s$^{-1}$ and red v=[9,29] km\,s$^{-1}$. For
SiO(2--1) the velocity ranges are: blue v=[-12,-1] km\,s$^{-1}$ and
red v=[6,15] km\,s$^{-1}$. The CO and SiO beams are shown at the
bottom left of each panel, respectively. Contour levels of the line
emission are always from 10 to 90\% (10\% steps) from the peak
intensities ($S_{\rm{peak\_blue}}(\rm{CO}(2-1))=50.8$\,Jy/beam,
$S_{\rm{peak\_red}}(\rm{CO}(2-1))=39.1$\,Jy/beam,
$S_{\rm{peak\_blue}}(\rm{SiO}(2-1))=1.3$\,Jy/beam,
$S_{\rm{peak\_red}}(\rm{SiO}(2-1))=1.1$\,Jy/beam). The line in the
left panel sketches the axis of the p-v diagram presented in Figure
\ref{pos_velo}. \label{19217_co}}
\end{figure}


\begin{figure}[htb] 
\caption{Continuum emission in IRAS\,20293+3952: the different
features are marked at the top right of each panel. At the bottom
left, we present the 1.3\,mm and 3\,mm beams, respectively. The mm data
are contoured from 10 to 90\% (10\% steps) from the peak emission
(Table \ref{continuum}). The cm observations are contoured from 40 to
90\% (10\% steps) from the peak emission of
0.8\,mJy/beam. \label{20293_cont}}
\end{figure}

\begin{figure}[htb] 
\caption{\footnotesize CO(2--1) emission in 20293+3952 (PdBI+PV): the
full and dotted/grey scale contours show the red- and blue-shifted emission,
respectively. The velocity ranges are: (a) blue [-40,-11]\,kms, red
[25,52]\,km\,s$^{-1}$; (b) blue [-10,0]\,km\,s$^{-1}$, red
[13,25]\,km\,s$^{-1}$; (c) blue [-40,0]\,kms, red
[13,52]\,km\,s$^{-1}$. The 1.3\,mm beam is shown at the bottom right
of each panel, the three stars and the square mark the positions of
the mm sources and the UCH{\sc ii} region, respectively. The arrows
and letters sketch the four outflows discussed in the main
body. Contour levels are always from 10 to 90\% (10\% steps) from the
peak intensities
($S_{\rm{peak\_blue\_ext}}(\rm{CO}(2-1))=10.0$\,Jy/beam,
$S_{\rm{peak\_red\_ext}}(\rm{CO}(2-1))=19.9$\,Jy/beam,
$S_{\rm{peak\_blue\_mod}}(\rm{CO}(2-1))=29.7$\,Jy/beam,
$S_{\rm{peak\_red\_mod}}(\rm{CO}(2-1))=13.0$\,Jy/beam,
$S_{\rm{peak\_blue\_all}}(\rm{CO}(2-1))=34.5$\,Jy/beam,
$S_{\rm{peak\_red\_all}}(\rm{CO}(2-1))=33.8$\,Jy/beam). \label{20293_co}}
\end{figure}

\begin{figure}[htb] 
\caption{SiO(2--1) in IRAS\,20293+3952: the
full and dotted contours show the red- and blue-shifted emission,
respectively. The 3\,mm SiO beam is presented at the bottom left, the three
stars and the square mark the positions of the mm sources and the
UCH{\sc ii} region, respectively. The arrows and letters sketch the
four outflows discussed in the main body. Contour
levels are always from 5 to 95\% (10\% steps) from the peak
intensities ($S_{\rm{peak\_blue}}(\rm{SiO}(2-1))=2.5$\,Jy/beam,
$S_{\rm{peak\_red}}(\rm{SiO}(2-1))=4.9$\,Jy/beam). \label{20293_sio}}
\end{figure}

\begin{figure}[htb] 
\caption{Red CO(2--1) emission in IRAS\,20293+3952: the grey-scale
shows the moderate velocities ([13,25]\,km\,s$^{-1}$) and the contours
present the high-velocity gas ([25,52]\,km\,s$^{-1}$). The 1.3\,mm
beam is shown at the bottom left, the three stars and the square mark
the positions of the mm sources and the UCH{\sc ii} region,
respectively. Contour levels are always from 10 to 90\% (10\% steps)
from the peak intensities as presented in
Fig. \ref{20293_co}. \label{red_20293}}
\end{figure}

\begin{figure}[htb] 
\caption{The grey-scale presents the H$_2$ emission (Stanke et al., in
prep.). The thick and thin black contours outline the red and blue CO
emission, respectively. The white contours present the 3.6\,cm VLA
observations of the UCH{\sc ii} region, and the three stars mark the
positions of the mm sources. The contouring is the same as in the
previous images. \label{20293_h2}}
\end{figure}

\begin{figure}[htb] 
\caption{Position-velocity diagrams of the outflows in
IRAS\,19217+1632 and IRAS\,20293+3952 (outflow A) presented in this
paper. Furthermore we show two position-velocity diagrams taken from
the literature: IRAS\,23033+5951 (observed with BIMA, Wyrowski et al.,
in prep.) and IRAS\,20126+4104 (observed with the IRAM 30\,m, Lebron
et al., in prep.). The horizontal lines mark the centers of the
outflows which always correspond to the main mm continuum
sources. Resolution elements are shown at the bottom left of each
panel.\label{pos_velo}}
\end{figure}

\end{document}